\UseRawInputEncoding
\documentclass[11pt,onecolumn,amssymb,nofootinbib]{revtex4}
\usepackage{amsmath, amsthm, amscd, amssymb}
\usepackage{graphicx, braket}
\usepackage{bm}
\usepackage{bbm}
\usepackage{epstopdf}

\begin{document}

\title{\bf Dynamical Gravastars} \bigskip

\author{Stephen L. Adler}
\email{adler@ias.edu} \affiliation{Institute for Advanced Study,
1 Einstein Drive, Princeton, NJ 08540, USA.}

\begin{abstract}
We combine the ideas of a Weyl scaling invariant dark energy action, which eliminates black hole horizons, with the ``gravastar'' idea of a jump in the hole interior from a normal matter equation of state to an equation of state where pressure plus density  approximately sum to zero.
Using the Tolman-Oppenheimer-Volkoff equation, which requires continuous pressure, we present Mathematica notebooks in which the structure of the gravastar is entirely governed by the action and the equation of state, with the radii where structural changes occur  emerging from the dynamics, rather than being specified in advance.  The notebooks work even with zero cosmological constant, but when the cosmological constant is nonzero, there is a very small black hole ``wind'' that we calculate by a relativistic extension of standard pressure driven isothermal stellar wind theory.
\end{abstract}

\maketitle

\section{Introduction}

\subsection{Mathematical black holes versus horizonless ``black'' holes such as ``gravastars''}

Extensive observations show that the universe contains a multitude of extremely compact objects, that are assumed to be mathematical black holes, as described in the monograph of Chandrasekhar \cite{chandra}. Mathematical black holes are solutions of the Einstein field equations characterized by just two parameters, the mass $M$ and the angular momentum per unit mass.  But the interpretation of astrophysical observations in terms of idealized mathematical black holes  has been questioned from several points of view.  In earlier papers reviewed in \cite{adler1}, including initially Adler and Ramazano\u glu \cite{AR}, and more recently followed up in Adler \cite{adler2} (with astrophysical applications in  \cite{adler3}, \cite{adler4})  we have proposed a novel Weyl scaling invariant form of the dark energy action, in which the integrand of the usual cosmological constant action contains an extra factor $g_{00}^{-2}$.  Hence this action is no longer interpretable as a ``vacuum energy''. Because of the factor $g_{00}^{-2}$, the Weyl scaling invariant action leads to vacuum ``black'' hole solutions\footnote{We use in this article the term ``black'' hole, with the quotes denoting an object which may have {\it no} event horizon, but which otherwise appears to astronomers very similar to the idealized mathematical black hole.  For mathematical black holes we continue to omit the quotes.}  with no event or apparent horizon, but with exterior  metrics outside the nominal horizon closely approximating the usual Schwarzschild or Kerr forms.   From a different perspective, several authors, as reviewed by Cardoso and Pani \cite{pani}, have proposed interior solutions for so-called ``exotic compact objects'' that appear black-hole like from the outside, but have no horizons and no interior singularity.  In particular, the ``gravastars'' proposed by Mazur and Mottola \cite{mazur} are based on assuming a discontinuous jump in the interior ``black'' hole equation of state, from a normal matter equation of state to the equation of state  proposed by Gliner \cite{gliner},  in which the pressure $p$ is minus the density $\rho$. Related ideas have been discussed via a condensed matter analogy in \cite{other}, \cite{other2}, \cite{khlopov1}, \cite{khlopov2}.

Our aim in the present paper is to combine the modified ``black'' hole ideas following from a Weyl scaling invariant dark energy action, with the proposal of a jump to a $p+\rho\simeq 0$ equation of state, to give a simple interior model of a modified ``black'' hole, in the spherically symmetric case.    The final result for our model takes the form of  Mathematica notebooks that are available online as supplementary material for this article \cite{notebook}.  Our model differs from that of Mazur and Mottola and the subsequent paper of Visser and Wiltshire \cite{visser} in
several significant respects.  First, we perform our entire analysis from the Tolman-Oppenheimer-Volkoff (TOV) equations for relativistic stellar structure, as augmented to include a Weyl scaling invariant cosmological constant action.  Second, we note that the TOV equations require that the pressure $p$ must be continuous,\footnote{For earlier work on gravastars with continuous pressure, but also continuous equation of state, see \cite{bened}.} whereas the energy density $\rho$ can have discontinuous jumps, so we implement the Gliner equation of state by a jump to negative energy density with positive pressure.  This of course violates the classical energy conditions, but from a semiclassical quantum matter point of view, the regularized energy density is known {\it not} to obey positivity conditions \cite{Wald}, \cite{Visser1} .  Third, we avoid assuming designated radii at which transitions take place. In our model, transitions follow dynamically from the equations of motion and the assumed equations of state, hence the title of this paper ``Dynamical Gravastars''.   And fourth, we smooth the jump in the equation of state by using a sigmoidal function in place of a Heaviside step function, so there are no exact discontinuities and accompanying surface densities to be considered. Thus we have a differential equation system that can be solved by the Mathematica  integrator NDSolve, which is powerful general tool for solving one dimensional differential equation systems, such as arise from our assumptions when restricted to spherical symmetry.

\subsection{Metric, gravitational action,  matter perfect fluid parameters, equation of state, and range of rescaled $\Lambda$ values}

The basic inputs to our model are the spherically symmetric metric, the action including the Einstein-Hilbert action and the dark energy action, and the assumed matter equation of state.
\begin{itemize}
\item {\bf Metric}  We write the static, spherically symmetric metric in the form
\begin{equation}\label{metric}
ds^2=e^{\nu(r)} dt^2-e^{\lambda(r)} dr^2 -  r^2 \big(d\theta^2+\sin^2(\theta)  d\phi^2\big)~~~,
\end{equation}
following the notation used in the monograph of Zeldovich and Novikov \cite{zeld} (except that we use geometrized units,  with the velocity of light $c$ and Newton's constant $G$ set equal to unity).

\item{\bf Gravitational action and matter parameters}
As in the papers reviewed in \cite{adler1}, we adopt  the postulate that the part of the gravitational action that depends only on the undifferentated metric $g_{\mu\nu}$, but involves no metric derivatives, is invariant under the Weyl scaling $g_{\mu\nu}\to \lambda g_{\mu\nu}$. Adoption of this postulate implies that the so-called  ``dark energy'' action has the  three-space general coordinate invariant,  but frame-dependent,  form
\begin{equation}\label{eff}
S_{\rm eff}=-\frac{\Lambda}{8 \pi } \int d^4x ({}^{(4)}g)^{1/2}(g_{00})^{-2}~~~,
\end{equation}
rather than the usually assumed vacuum energy form
\begin{equation}\label{vacen}
S_{\rm energy}=-\frac{\Lambda}{8 \pi } \int d^4x ({}^{(4)}g)^{1/2}~~~,
\end{equation}
where $\Lambda$ is the observed cosmological constant, and ${}^{(4)}g=-\det(g_{\mu\nu})$.
Since the unperturbed Friedmann-Lema\^itre-Robertson-Walker (FLRW) cosmological metric has  $g_{00}=1$, in this context the action of Eq. \eqref{eff} mimics the standard cosmological constant action of  Eq.  \eqref{vacen}, but when $g_{00}$ deviates from unity, their consequences differ. There are a number of motivations, which are reviewed in detail in \cite{adler1}, for studying the possibility that dark energy arises from the action of Eq. \eqref{eff}. Here,  suffice it to say that assuming that dark energy arises as a vacuum energy from the action of Eq. \eqref{vacen} leads to the cosmological constant fine tuning problem, which is not implied by alternative forms of the dark energy action, such as Eq. \eqref{eff}.

To this dark energy action we add the standard Einstein-Hilbert gravitational action constructed from derivatives of the metric,
\begin{equation}\label{grav}
S_g=\frac{1}{16 \pi} \int d^4x ({}^{(4)}g)^{1/2} R~~~,
\end{equation}
with $R$ the curvature scalar.  Finally, we include a matter action $S_m$ to respresent material that is inside the ``black'' hole, which we assume takes the form of a relativistic perfect fluid.  The gravitational field equations are obtained by varying the the sum $S_{\rm eff}+S_g+S_m$ with respect to the spatial components $g_{ij}$ of the metric, and then imposing covariant conservation (or equivalently, Bianchi identities for the metric) to infer the remaining components, a procedure discussed in detail in \cite{adler1} and \cite{AR}. The result is that the total pressure and energy density $\hat p$ and $\hat \rho$, including contributions from the dark energy action of Eq. \eqref{eff},  are related to the matter pressure and energy density $p$ and $\rho$, by (with $\kappa \equiv  8 \pi$)
\begin{align}\label{totals}
\hat p=& p - \frac{\Lambda}{\kappa} e^{-2\nu(r)}~~~,\cr
\hat \rho =& \rho- \frac{3\Lambda}{\kappa} e^{-2\nu(r)}~~~,\cr
\end{align}
which obey
\begin{align}\label{sumdiff}
\hat \rho-3\hat p= &\rho-3p~~~,\cr
\hat \rho+\hat p =& \rho+p-\frac{4\Lambda}{\kappa} e^{-2\nu(r)}~~~.\cr
\end{align}

\item {\bf Equation of state}  We assume the following equation of state for the matter content of the model.  For  pressure $p$ less than a critical value ``pjump'' the matter obeys a relativistic equation of state  $\rho=3 p$, which by Eq. \eqref{sumdiff} implies $\hat \rho=3 \hat p$.  For pressure greater than pjump, we assume that the matter jumps to an  equation of state $p+\rho=\beta$, which has the Gliner form as as modified by addition of a  ``bag constant'' $\beta$.  This addition  plays a role similar to that played by the non-isotropic pressure introduced by  Cattoen et al. \cite{cattoen} and others.\footnote{For the values  $\beta=.1,~ .01 $ studied in our numerical examples, the cosmological constant can be set to zero without visibly changing the plotted results, so a jump triggered by the value of $\hat p$ is equivalent to one triggered by the value of $p$.  For $\beta=.001$ the cosmological constant has a small effect on the numerical output, suggesting that for extremely small $\beta$ values there could be a substantive difference between a jump triggered by $\hat p$ and one triggered by $p$.  This question remains to be studied in future work.}

\item{\bf  Range of rescaled $\Lambda$ values}  The above ingredients are the content of our model.   In programming the model, it is convenient to rescale to dimensionless variables for which the matter pressure at the center of the modified ``black'' hole is unity, $ p(0)=1$. After this rescaling, discussed in detail in Appendix B and shown in Table II for a $10^6 M_\odot$ hole,  the rescaled parameter lambda is very small. For the numerical examples corresponding to $\beta=.1, ~.01,~  .001$ given below, lambda has a small effect on the displayed graphs, and could be set equal to zero.  This shows that when matter is present,  an interior jump to the Gliner equation of state suffices to eliminate the horizon, without needing the presence of the Weyl scaling invariant dark energy action which was used \cite{AR}  to eliminate the horizon in the vacuum Einstein equation case.  However, the calculation of the black hole wind given later depends crucially on $\Lambda$ having a nonzero positive value
\end{itemize}

\subsection{The modified TOV equations, continuity conditions, and initial conditions}

The standard way of computing the structure of relativistic stars is through the TOV equations, which combine the Einstein equations for the metric coefficients $\nu(r)$ and $\lambda(r)$ with the covariant conservation equations for the matter content of the star.  A succinct derivation is given in \cite{zeld} and a pedagogical exposition is given in the monograph of Camenzind \cite{camen}.  In terms of the total pressure and energy density $\hat p$ and $\hat \rho$, the modified TOV equations are\footnote{The pressure equation is often referred to in the singular as ``the TOV equation''.}
\begin{align}\label{TOV}
\frac{d\hat m(r)}{dr}=&4 \pi  r^2\hat \rho(r)~~~,\cr
e^{-\lambda(r)}=&1-\frac{2\hat m(r)}{r}~~~,\cr
\frac{d\nu(r)}{dr}=&\frac{\hat N_\nu}{1-2\hat m(r)/r}~~~,\cr
\hat N_\nu=&(2/r^2)(\hat m+4  \pi r^3 \hat p)~~~,\cr
\frac{d\hat p}{dr}=&-\frac{\hat \rho+\hat p}{2}\frac{d\nu(r)}{dr}~~~.\cr
\end{align}
The final equation, for $d\hat p/dr$, can be converted to an equation for $dp/dr$ by using Eqs. \eqref{totals} and \eqref{sumdiff},
\begin{equation}\label{convert}
\frac{dp}{dr}=\frac{d\hat p}{dr}-2 \frac{d\nu}{dr} \frac{\Lambda}{\kappa} e^{-2\nu(r)}=-\frac{\hat \rho+\hat p}{2}\frac{d\nu(r)}{dr}-2 \frac{d\nu}{dr} \frac{\Lambda}{\kappa} e^{-2\nu(r)}=-\frac{\rho+ p}{2}\frac{d\nu(r)}{dr}~~~,
\end{equation}
showing that $dp/dr$ vanishes when $\rho+p$ vanishes, as in the postulated Gliner equation of state.

Assuming that all quantities appearing on the right hand side of the TOV equations are bounded, the one dimensional version of the standard ``pillbox'' argument implies that $\hat m(r)$, $\nu(r)$, $\hat p(r)$, and $p(r)$ must all be continuous functions of $r$, with no jump
discontinuities. (See Appendix A.) However, $\hat \rho(r)$ and $\rho(r)$ can have finite jump discontinuities, since $\hat \rho(r)$ only appears on the right hand side of the TOV equations.

The initial value conditions for the TOV equations can be taken as $ p(0)=1$, $\hat m(0)=0$, and $\nu(0)=$``nuinit'', where nuinit is fixed {\it a posteriori} by requiring a match to the Schwarzschild metric value $\nu(\infty)=0$ at asymptotically large $r$.

\subsection{Exterior space limit}

From Eq. \eqref{TOV}, we see that when    $\hat p=0$ and $\hat \rho=0$, we have\footnote{More generally, Eq. \eqref{ddnom} holds when $ \hat p=- \hat\rho\neq 0$.  Using this, one finds that when $\beta=\Lambda=0$, the interior solution for $r$  below the jump is given exactly by $p(r)=1$, $\rho(r)=-1$, $m(r)=-4\pi r^3/3$, $\hat N_\nu=16 \pi r/3$, and $\nu(r)=\nu(0)+ \log\big(1-2 m(r)/r\big)$, with continuity of $p$ requiring ${\rm pjump}=1$.}
\begin{equation}\label{ddnom}
\frac{d}{dr}(1-2 \hat m/r)=(2/r^2) \hat m -(2/r)4 \pi r^2  \hat \rho= (2/r^2) (\hat m+ 4 \pi r^3 \hat p)=\hat N_\nu~~~.
\end{equation}
Hence in the limit $\hat p=\hat \rho=0$, the differential equation for $\nu(r)$ becomes
\begin{equation}\label{dnu}
\frac{d\nu(r)}{dr}=\frac{ \frac{d}{dr}(1-2 \hat m/r)   }{1-2\hat m(r)/r}=\frac{d}{dr}\log(1-2 \hat m/r)~~~,
\end{equation}
which, with the asymptotic boundary condition $\nu(\infty)=0$, integrates to
\begin{align}\label{nur}
\nu(r) = &\log(1-2 \hat m/r)~~~,\cr
e^{\nu(r)}=&1-2 \hat m/r~~~.\cr
\end{align}
Similarly, from Eq. \eqref{TOV} we see directly that
\begin{equation}\label{lambdar}
e^{\lambda(r)}=1/(1-2 \hat m/r)~~~.
\end{equation}
So as $\hat p$ and $\hat \rho$ approach zero, the solution to the TOV equations approaches the free space Schwarzschild solution corresponding to mass $\hat m(r=\infty)$.  We shall see this behavior in the Mathematica notebooks given below, when the initial value $\nu(0)={\rm nuinit}$ is fixed to guarantee that $\nu(\infty)=0$.

\subsection{Sigmoidal ``theta'' and ``delta'' functions to smooth the equation of state jump}

Although the TOV equations allow the energy density $\rho$ to have a finite jump discontinuity, it is convenient in solving these equations numerically to smooth this jump, by using a sigmoidal version of the standard Heaviside step function $\theta(x)$.  We do this by defining
\begin{align}\label{smooththeta}
\theta_\epsilon(x)=& \frac{1}{1+e^{-x/\epsilon}}~~~,\cr
\theta_\epsilon(-x)=&\frac{1}{1+e^{x/\epsilon}}=e^{-x/\epsilon}\theta_\epsilon(x)~~~,\cr
\end{align}
with $\epsilon>0$ very small.  The corresponding smoothed extension of the standard Dirac delta function $\delta(x)$ is
\begin{equation}\label{smoothdelta}
\delta_\epsilon(x)=\frac{d\theta_\epsilon(x)}{dx}=\frac{1}{\epsilon} \theta_\epsilon(x) \theta_\epsilon(-x)~~~.
\end{equation}
We shall use both of these smoothed functions in the programming.

\section{Notebook for the model}

Sample Mathematics notebooks for our model, for $\beta$ parameter values $\beta=.1$, $\beta=.01$, and $\beta=.001$, can be downloaded at the  URL given in \cite{notebook}. These notebooks were written using Mathematica version 12.2, but should work in most earlier versions \cite{notebook}.
   The programs begin with a list of numerical parameters, as shown for the three $\beta$ values in Table I.
\begin{table} [ht]
\caption{Numerical parameters for the Mathematica notebooks.  For $\beta=.1$ and $.01$ the program is not sensitive to the values of lambda shown in Tables I and II, and gives the same graphs for lambda of 0.   For $\beta=.001$, we could only get a good asymptotic match for lambda of $10^{-44}$ and smaller. Thus the desired value of $.4 \times  10^{-42} $ was not attainable, and we used $10^{-44}$.  In the TOV.001 notebook, to change lambda to $0$ the value of nuinit  should be changed to $-50.60$.}
\centering
\begin{tabular}{c  c c c }
\hline\hline
 notebook name&~~TOV.1~~&~~TOV.01~~ & ~~TOV.001~~ \\
\hline
beta     &   $.1$    &  $.01$     & $.001$   \\
nuinit     &   $-14.70$    &   $-21.255$    &  $-50.75$  \\
pjump     &    $.7 $  &  $.95 $    &  $.98 $ \\
lambda    & $.3 \times  10^{-34} $    &   $ 10^{-36} $    & $.4 \times  10^{-42},\, {\rm used}\, 10^{-44} $   \\
rmax     &   $10$    &  $60$     & $80,000$   \\
\hline
rmin     &   $10^{-7}$   &  $10^{-7}$      & $10^{-7}$    \\
alpha0     &    $-1$   &    $-1$    &   $-1$  \\
alpha1     &   $3$    &   $3$       &   $3$    \\
kappa     &   $8\pi$    &   $8\pi$       &    $8\pi$   \\
kappa2     &   $4\pi$       &  $4\pi$      &  $4\pi$   \\
eps     &   $.001$    &    $.001$      &  $.001$     \\
\hline\hline
\end{tabular}
\label{tab1}
\end{table}

Following the initial parameter values list, there are five function definitions.  The sigmoidal function of Eq. \eqref{smooththeta}  is implemented by theta[x\_]:=1/(1+Exp[-x/eps]), while the equations of Eq. \eqref{totals} which construct $\hat p$ and $\hat \rho$ are implemented by phat[x\_,y\_]:=x-(lambda/kappa)\textasteriskcentered Exp[-2\textasteriskcentered{}y] and rhohat[x\_,y\_]:=rho[x]-(3\textasteriskcentered{}lambda/kappa)\textasteriskcentered{} Exp[-2\textasteriskcentered{}y].  Finally, the switch in the equation of state is implemented by the functions alphas[x\_]:=alpha0\textasteriskcentered{}theta[x-pjump]+alpha1\textasteriskcentered{}theta[pjump-x]   and  rho[x\_]:=alphas[x]\textasteriskcentered{}x +beta\textasteriskcentered{}theta[x-pjump].   In using these functions in the differential equation solver, x will always be $p[r]$ and y will always be $\nu[r]$.

After the function definitions, there follows setup of the system of differential equations  to be solved.  The variables  nu[r],  p[r], and emhat[r] correspond to $\nu(r)$, $p(r)$, and $\hat m(r)$ in the TOV equations of Eq. \eqref{TOV}, and have respective initial values nuinit, 1, and 0 respectively, given in the first three lines within ``system=\{....\}''.    The second three lines are the TOV differential equations, constructed using the functions defined  in the preceding paragraph.   The remainder of the notebook consists of the command NDSolve for the system of equations, extraction of the solution from the interpolating functions constructed by NDSolve, and computation of certain auxiliary quantities together with  graphical plotting.  The integration range is taken to start from  $r=10^{-7}$ rather than  $r=0$ to avoid zero divides; the maximum $r$ value needed for the integration range and plots depends on the value of $\beta$.

\section{Some sample output}

The three notebooks TOV.1,  TOV.01, and TOV.001 correspond respectively to choices $\beta=.1$, $\beta=.01$, and $\beta=.001$ in the inner region equation of state $p+\rho=\beta$.  If $\beta$ were taken as zero, the interior pressure would not evolve from its initial value $p(0)=1$, so for generality we have taken a nonzero value of $\beta$.\footnote{A nonisotropic pressure term \cite{cattoen} would have a similar effect.}    But the chosen values may not be representative of realistic ``black'' hole solutions, which may correspond to much smaller $\beta$ values. These will be hard to implement in our Mathematica notebooks because some of the computed quantities, such as $\nu$,  will become very large.  To explore a full range of $\beta$ values, it is important to try to develop analytic approximations to the TOV equation solutions.

The parameter values pjump in the notebooks, where the equation of state jumps as a function of pressure $p$,  represent arbitrary choices, not reflecting any attempt at a systematic survey.  We expect some quantitative features of the numerical output to depend strongly on where this jump is placed.  So the results presented in Table II and in the Figures should be considered as a sampling of the solution space.

A key feature of the numerical solution is that once nuinit is adjusted to give a match to a Schwarzschild solution at large $r$, the rest of the solution is determined by the dynamical equations and the assumed equations of state.  In Table II we give, as computed in the three notebooks,
the approximate rescaled hole mass $M$, the rescaled cosmological constant lambda for a $10^6 M_\odot$ hole (see Appendix B),  and an auxiliary quantity that sets the scale for the black hole wind when multiplied by $3\Lambda/\kappa$.
\begin{table} [ht]
\caption{Numerical parameters derived from the output of the  Mathematica notebooks.  In the TOV.001 notebook we used lambda of $10^{-44}$ since the target of $.4 \times 10^{-42}$ was not attainable. }
\centering
\begin{tabular}{c  c c c }
\hline\hline
 notebook name&~~TOV.1~~&~~TOV.01~~ & ~~TOV.001~~ \\
\hline
${\rm rescaled~hole~mass}~M$  &   $ 3.03  $   &  $ 16.5  $     & $27600   $   \\
lambda for $10^6 M_\odot$ hole  &   $ .3 \times 10^{-34}  $   &  $ 10^{-36}  $     & $ {\rm used}\, 10^{-44}   $   \\
$\exp\big(-2\nu(3M)\big)-1$    &   $ 7.86$    &   $  7.91 $    &  $ 7.82  $  \\
\hline\hline
\end{tabular}
\label{tab2}
\end{table}

Since the qualitative features of the three notebooks are very similar, we give in the first eight Figures only plots for the $\beta=.01$ notebook.  In Fig. 1, we plot the TOV denominator ${\rm denom}={\cal D}=1-2 \hat m/r$, which becomes very small at the nominal hole radius $2M$, but never vanishes. The kink at $r=28.5$ corresponds to the equation of state jump (see Fig. 8), where $\hat m$ starts to increase from negative values.  The kink at $r=33 \simeq 2M$, which on a finer scale can be seen to be smooth, and not a cusp, corresponds to the merger into an exterior Schwarzschild solution where $\hat \rho$ and $\hat p$ vanish.  In Fig. 2, we show $\hat m(r)=1-{\cal D}(r)*r/2$, which gives a determination of the effective hole mass $M$ from the metric coefficient $\lambda(r)$ (not to be confused with lambda, the Mathematica notebook label for the rescaled cosmological constant $\Lambda$).   In Fig. 3 we plot  $\big(1-\exp(\nu(r)\big)*r/2=  M(r)$, giving a determination of the effective hole mass $M$ from the metric coefficient $\nu(r)$.  Achieving a leveling off of the slope on the right of this plot was used to tune the initial value nuinit, since  this slope just measures $\big(1-\exp(\nu(\infty)\big)/2$, and so a vanishing slope corresponds to
the desired condition $\nu(\infty)=0$.     Increasing  nuinit from the optimal value results in the right hand flat portion of the plot tilting downwards, and decreasing  nuinit from the optimal value results in the right hand flat portion of the plot tilting upwards.  In Fig. 4 we plot $\hat p \simeq p$, which shows that it is a positive monotonically decreasing function of $r$, which vanishes rapidly above $2M$.  In Fig. 5 we give the corresponding plot of $\hat \rho \simeq \rho$, with the equation of state jump clearly visible, as well as the rapid vanishing above $2M$.   In Fig. 6 we plot $\nu(r)$, and in Fig. 7 we plot the quantity $(3/\kappa) \exp\big(-2\nu(r)\big)$.  Finally, in Fig. 8 we plot $\hat \rho(r)/\hat p(r) \simeq \rho(r)/p(r)$, again clearly showing the equation of state jump at $r=28.5$.
Some other graphs of interest are given in the notebooks, and the reader who downloads the notebooks can readily  plot others.

\section{Stability analysis}

Analyzing stability of  relativistic star interior solutions obtained from the TOV equations can be done by a method developed by Chandrasekhar \cite{chandra1} and reviewed in  \cite{bard} and \cite{hanss}.  Starting from the eigenequation for time-dependent normal modes around the TOV static solution, one constructs a Rayleigh-Ritz variational principle for the eigenvalues $\omega^2$ and eigenfunctions $u(r)$,
\begin{equation}\label{rrprin}
\omega^2=\frac{\int_0^R dr [P (du/dr)^2-Q u^2]}{\int_0^R dr W u^2 }~~~,
\end{equation}
where $R$ is the radius of the star and $u$ is a trial eigenfunction.   The functions $P(r),\,Q(r),\,W(r)$ are constructed
from the metric coefficients and the interior equation of state according to
\begin{align}\label{PQW}
P=&\exp\big((\lambda+3 \nu)/2\big)r^{-2}\gamma p~~~,\cr
Q=&-4\exp\big((\lambda+3 \nu)/2\big)r^{-3} dp/dr-8\pi \exp\big(3(\lambda+ \nu)/2\big)r^{-2}p(p+\rho)\cr
+&\exp\big((\lambda+3 \nu)/2\big)r^{-2}(p+\rho)^{-1} (dp/dr)^2~~~,\cr
W=&\exp\big((3\lambda+ \nu)/2\big)r^{-2}(p+\rho)~~~,\cr
\end{align}
with $\gamma$ the ``adiabatic index''
\begin{equation}\label{index}
\gamma=(p+\rho)p^{-1} (\partial p/\partial \rho)|_{\rm constant~entropy}~~~.
\end{equation}
In applying this recipe, we rewrote Eq. \eqref{index} as
\begin{equation}\label{index1}
\gamma p=(p+\rho)/(d \rho/d p)~~~,
\end{equation}
and since $(d \rho/d p)$ has a zero near pjump, we rewrote $(d \rho/d p)^{-1}$ as a principal value
\begin{equation}\label{prinval}
(d \rho/d p)^{-1}=  \lim_{\epsilon_1 \to 0} \frac{(d \rho/d p)}{(d \rho/d p)^2+ \epsilon_1^2}~~~.
\end{equation}
We took the trial function as $u(r)=r^3 (r-R)^2$.  The factor $r^3$ is needed to satisfy the boundary condition stated in Eq. (7a) of \cite{bard}.  The boundary condition of Eq. (7b) of \cite{bard} requires the vanishing of $-\exp(\nu/2)r^{-2}\gamma p du/dr$  at the surface of the hole.  According to Eq. \eqref{index1}, Eq. \eqref{sumdiff},  and Fig. 9,
\begin{equation}\label{outer}
\gamma p|_{2M} = [(\hat p+ \hat \rho)|_{2M} +(4\Lambda/\kappa)\exp(-2\nu(2M))]/3\neq 0~~~,
\end{equation}
so $du/dr$ must vanish at $R=2M$, requiring the factor of $(r-R)^2$. (The nonvanishing of $p + \rho$ at the hole surface will also play a key role in the wind caculation of the next section.)
Evaluating the integral in Eq. \eqref{rrprin} with $R=2M\simeq 6.06 $ in the $\beta=.1$ computation, and with choices of $\epsilon_1=.1,\,.01,\,.001$ in the principal value construction of Eq. \eqref{prinval},   gives $\omega^2=.002>0$, compatible with stability.\footnote{The sequence of $R$ values $6.0610,\,6.0608,\,6.0606,\,6.0605$ gives  the respective results $.0019200,\,.0019202,\,.0019204,\,.0019206$, whereas the $R$ value $6.0604$ gives a warning of slow convergence of the numerator integral.}

However, two caveats are in order.  The first caveat is that since the right hand side of Eq. \eqref{outer} is very small (but nonzero),
 it is reasonable to ask what happens if it is approximated by zero.  Then one need not require $du/dr|_{2M}=0$, allowing a trial function $u(r)=r^3$.  For this trial function one finds $\omega^2 =-.007<0$, corresponding to instability.  So the issue of the outer boundary condition is clearly subtle.  The second caveat is that having a principal value singularity in the integral for the stability test is not anticipated in the standard applications of this test, or in the Sturm-Liouville theory on which this test is based.   Thus we regard the issue of stability or instability of our model as not definitive; further study of the case when the pressure is continuous, but the energy density has a jump, is needed.

\section{Relativistic calculation of the black hole wind}

Rewriting  Eq. \eqref{totals} as
\begin{align}\label{totals1}
 p=& \hat p +\frac{\Lambda}{\kappa} e^{-2\nu(r)}~~~,\cr
  \rho =& \hat \rho +\frac{3\Lambda}{\kappa} e^{-2\nu(r)}~~~,\cr
\end{align}
and using the fact that $\hat p$ and $\hat \rho$ vanish in the exterior region, we see that the matter pressure $p$ and energy density $\rho$ are nonvanishing in the exterior. This brings into play the mechanism for an isothermal pressure driven wind pioneered by Parker \cite{parker}.  In the Parker calculation, one combines the equations for gas momentum conservation and energy conservation in the presence of the gravitational field of a star of mass $M$, with the gas equation of state $p=a^2 \rho$, to get an equation for the gas velocity $V$ of the form
\begin{equation}\label{parkereq}
\frac{1}{V} \frac{dV}{dr}=\left(\frac{2a^2}{r}-\frac{M}{r^2}\right)/\Big(V^2-a^2\Big)~~~.
\end{equation}
The numerator of this equation vanishes at the {\it critical distance} $r_c=M/2a^2$, and the only solution of Eq. \eqref{parkereq} for which the velicity gradient is positive at all distances $r$ is one for which $ V(r_c)=a$, defining the {\it critical solution}.  From the properties of the critical solution, and the radially conserved flux per steradian
\begin{equation}\label{flux1}
 F=  r^2 \rho(r) V(r)~~~,
\end{equation}
one calculates the wind rate of mass loss from the star.  For a very clear pedagogical discussion of the Parker mechanism, see \cite{lamers}.

The above formulas are all nonrelativistic as appropriate to a low velocity gas acted on by Newtonian gravity.  To discuss the wind emanating from our gravastar model, general relativistic extensions are needed.  For the equation of state, we continue to write $p=a^2 \rho$, with $a=1/\surd{3}$ for a gas of relativistic particles.  The energy and momentum conservation equations are obtained from the covariant conservation equations for the energy-momentum tensor, describing a relativistic gas with radial velocity $V(r)$ in the presence of the  general spherical metric $g_{\mu\nu}$ of Eq. \eqref{metric}.  This energy momentum tensor takes the perfect gas form
\begin{equation}\label{enmomgas}
T^{\mu\nu}=[p(r)+\rho(r)] U^{\mu}U^{\nu} - p(r) g^{\mu\nu}= (1+a^2) \rho(r) U^{\mu}U^{\nu}   -a^2 \rho(r) g^{\mu\nu}~~~,
\end{equation}
with the four-velocity $U^{\mu}(r)$ given by
\begin{align}\label{fourvel}
U ^{\mu}=&U^0\big(1,V(r),0,0\big)~~~,\cr
U^0=&1/[\exp\big(\nu(r)\big)-\exp\big(\lambda(r)\big) V(r)^2 ]^{1/2}~~~,\cr
1=&g_{\mu\nu} U^{\mu}U^{\nu}~~~.\cr
\end{align}

From this point on the algebra gets complicated.  We use Mathematica to form the covariant divergence
\begin{equation}\label{ddeff}
D^{\nu}= \nabla_{\mu}T^{\mu\nu}~~~,
\end{equation}
 giving the conservation equations $D^{0}=0$ and $D^{r}=0$.  We found it convenient to use the linear combinations $D^{r}-V(r)D^{0}=0~,\,D^{0}=0$ in the next step, where we use Mathematica to solve for
$\rho^\prime(r)/\rho(r)$ and $V^\prime(r)/V(r)$, giving
\begin{align}\label{simeqresults}
\frac{\rho^\prime(r)}{\rho(r)}=&\frac{N_\rho}{[V(r)^2  \exp\big(\lambda(r)\big)-a^2 \exp\big(\nu(r)\big)]}~~~,\cr
\frac{V^\prime(r)}{V(r)}=&\frac{N_V}{[V(r)^2  \exp\big(\lambda(r)\big)-a^2 \exp\big(\nu(r)\big)]}~~~.\cr
N_\rho=&(1+a^2) \big[-4 \exp\big(\lambda(r)\big)V(r)^2+\exp\big(\nu(r)\big)r \nu^\prime(r) -\exp\big(\lambda(r)\big)r V(r)^2 \nu^\prime(r)\big]/(2r)~~~,\cr
N_V=&-\big[-4 a^2 \exp\big(\nu(r)\big)+4 a^2 \exp\big(\lambda(r)\big)V(r)^2-a^2 \exp\big(\nu(r)\big)r \lambda^\prime(r) +\exp\big(\lambda(r)\big)r V(r)^2 \lambda^\prime(r)\cr+&\exp\big(\nu(r)\big)r \nu^\prime(r) -2 \exp\big(\lambda(r)\big)r V(r)^2 \nu^\prime(r) +a^2 \exp\big(\lambda(r)\big) r V(r)^2 \nu^\prime(r)\big]/(2r)~~~.\cr
\end{align}

The denominators in the above equations are the relativistic generalization of that in Eq. \eqref{parkereq}, and so the critical solution is defined now by
\begin{equation}\label{relcrit}
V(r)^2=a^2 \exp\big(\nu(r)-\lambda(r)\big)~~~.
\end{equation}
Substituting this into $N_V$ and simplifying, we get
\begin{equation}\label{rceq}
N_V|_{\rm critical~ solution}= (1-a^2)\exp\big(\nu(r)\big)[4 a^2 +(a^2-1)r \nu^\prime(r)]/2r~~~,
\end{equation}
the vanishing of which determines the critical radius to be the solution of
\begin{equation}\label{rcritrel}
4 a^2 +(a^2-1)r \nu^\prime(r)=0~~~.
\end{equation}

In the relativistic case when $a^2$ and $\nu(r)$ are not small, we can proceed by observing that in the exterior region $\exp\big(\nu(r)\big)$ is very closely approximated by $1-2M/r$, as shown in Fig. 9 in the $\beta=.1$ computation.  Thus we can approximate
\begin{align}\label{nuapprox}
\nu \simeq& \log(1-2M/r)~~~,\cr
r \nu^\prime \simeq& 2M/(r-2M)~~~,\cr
\end{align}
which when substituted into Eq. \eqref{rcritrel} gives
\begin{equation}\label{rcritsoln}
r_c=\frac{2M(1+3a^2)}{4a^2}~~~.
\end{equation}
For $a^2=1/3$, and any hole mass $M$, this gives $r_c=3M$, which is just the photon sphere radius \cite{photon}, the boundary between black hole photon orbits that spiral out to infinity, and ones that fall into the hole.

To recover the nonrelativistic calculation, we treat $a^2$ and $M/r$  as small relative to 1, and take $\exp(\nu) \simeq 1$,  $r\nu^\prime \simeq 2M/r$.
Then Eqs. \eqref{simeqresults} -- \eqref{rceq} reduce to
\begin{equation}\label{nonrellim}
\frac{V^\prime(r)}{V(r)} \simeq \left[\frac{2a^2}{r}-\frac{M}{r^2}\right]/\Big[V(r)^2-a^2\Big]~~~,
\end{equation}
which agrees with Eq. \eqref{parkereq}.

From Eq. \eqref{nuapprox}, we also understand the second line in Table II, which gives $ \exp\big(-2\nu(3M)\big)-1\simeq 7.82~{\rm to}~7.91$ for the $\beta =.1,\, .01, \,.001$ calculations.   We have
\begin{equation}\label{expfactor}
\exp\big(-2\nu(3M)\big)-1 \simeq 1/(1-2/3)^2-1 =8~~~.
\end{equation}

The remaining step to compute the wind magnitude is to identify the relativistic analog of the nonrelativistic conserved flux per steradian
$ r^2 \rho(r) V(r)$.  As shown in Appendix C, when $p(r)=a^2 \rho(r)$, this is given by the formula
\begin{align}\label{consflux}
 F=&  (1+a^2) \exp\big((\lambda(r) + 3 \nu(r))/2\big) r^2 \rho(r) V(r)/[\exp\big(\nu(r)\big)- \exp\big(\lambda(r)\big)V(r)^2 ]~~~,\cr
dF/dr=&0~~~.\cr
\end{align}
We evaluate this expression at $r=3M$, using Eq. \eqref{relcrit} to get $V^2(3M)$,  using $\lambda(3M)\simeq -\nu(3M)$, $\exp\big(\nu(3M)\big)\simeq 1/3$, and $\exp\big(-2\nu(3M)\big)\simeq 9$.  For the net density driving the wind, we substitute in Eq. \eqref{consflux} the difference between the density value at $r=3M$ and the density value at $r=\infty$, that is    $\rho(3M)\to \frac{3\Lambda}{\kappa}[ e^{-2\nu(3M)}-1]$,    giving for the mass loss rate $\dot M$ from the wind
\begin{equation}\label{massloss}
\dot M= 4 \pi F = 24 \surd{3} \Lambda M^2~~~.
\end{equation}
Since in geometrized units the mass $M$ has dimensions of length, the combination $\Lambda M^2$ is invariant under the scaling of Appendix B, so Eq. \eqref{massloss} applies directly to the physical cosmological constant and hole mass.  For a solar mass hole, this gives an evaporation rate $\dot M_\odot/M_\odot \sim .76 \times 10^{-31} {\rm year}^{-1}$, much larger than the Hawking radiation evaporation rate \cite{hawk}  of $\dot M_\odot/M_\odot \sim .5 \times 10^{-67} {\rm year}^{-1}$.
However, the rate given in Eq. \eqref{massloss}, which corresponds to a non-accreting hole,   is much too small to play a role in astrophysical processes such as galaxy formation.  To address astrophysical implications of horizonless holes, accretion of infalling matter will have to be taken into account.

\section{Suggested extensions of the calculations}

We give several suggestions for extension of the calculations of the preceding sections.

\begin{itemize}
\item  {\bf Adjustment of nuinit}   Adjustment of the initial value $\nu(0)={\rm nuinit}$ to achieve $\nu(\infty)=0$, as needed to give a match to a Schwarzschild solution at spatial infinity, is tedious.  We have constructed the Mathematica notebooks with simplicity in mind, but they could  be enhanced by adding an overall iterative loop to automatically adjust nuinit to achieve a flat $M(r)$ for large $r$.

\item  {\bf Exploring the parameter space} There are three parameters, $\beta, \, \Lambda,$ and pjump, and it would be of interest to explore the behavior of the model over a wide range of these.  This will likely push the capabilities of the Mathematica integrator.  It may require development of semi-analytic approximation methods to do this, including a semi-analytic model for the dip in the denominator ${\cal D}$ near $r=2M$.  A related question is the failure of NDSolve to give results when $\Lambda \propto {\rm lambda} $ is too large, as we noted in the $\beta=.001$ notebook.  This is present at larger lambda values in the other notebooks as well, and we have not determined the source of this breakdown.

\item {\bf Axially symmetric, rotating extension}  Just as Kerr black holes are the natural extension of spherically symmetric Schwarzschild black holes, there should be an axially symmetric rotating extension of the dynamical gravastars calculated in this paper.  This will require solving two dimensional, as opposed to one dimensional  differential equations, which can be considerably more difficult.  For a discussion of technical difficulties encountered in an attempt to extend the results of the free space analysis of \cite{AR} to the axial case, see the final sections and Appendices of \cite{adler2}.

\item {\bf Accreting holes}  A first step towards analysis of accreting ``black'' holes will be to do a systematic study of orbits of incoming particles, using the metric calculated in the Mathematica notebooks.  For large impact parameters compared with the hole radius, the orbits will be similar to conventional black hole orbits, but for impact parameters of order a few times the hole radius and smaller, there will be systematic changes.  In Fig. 10 we have plotted $dr/dt$ for a radially infalling particle from the TOV.001 notebook.  This plot only takes account of gravitational forces, using the equation \cite{weinvel}
    \begin{equation}\label{infallgraph}
    \left(\frac{dr}{dt}\right)^2= \exp\big(\nu(r)-\lambda(r)\big)[1- \exp\big(\nu(r)\big)]~~~.
    \end{equation}
    This plot gives an upper bound on $dr/dt$ when the angular momentum per unit mass is nonzero.
     Evidently there is a small region near the center of the hole where the particle velocity is so small that a particle may be effectively trapped, but there is a larger outer region where a particle may escape back to infinity on a physically relevant timescale.  This could give a plausible mechanism for black holes to nucleate galaxy formation \cite{adler4}  by re-emitting or ``leaking'' most accreting particles as a sizable black hole wind, while retaining a small fraction of accreting particles which contribute to simultaneous black hole growth.

\item {\bf Dependence of results on equation of state} One could reconfigure the Mathematica notebooks to study other types of equation of state  in place of the ones used in the notebooks. For example, one could look at the widely used polytropic equation of state $p=K \rho^\Gamma$ in the exterior region, and alternatives to the Gliner equation of state in the interior region.

\item {\bf Stability}  We gave preliminary results concerning stability of our solutions, but not an analysis that we consider conclusive.  This will require extension of the current methods for stability analysis to the case when the energy density is not required to be positive.

\item {\bf Models with pressure jump}  Since we have smoothed the discontinuities with a sigmoidal function, one could use this method in the TOV equations context to study models in which the energy density remains positive and the pressure jumps, as in the original gravastar papers.  Can dynamical models, with no preset radii for transitions, be obtained this way?  We did not invest much time pursuing this, because the models with an energy jump work so well, and are natural to the structure of the TOV equations, but it is worth investigating.

\end{itemize}

\section{Possible Observational Effects}

If what were thought to be mathematical black holes are really some type of exotic compact object, there will be consequences for observational astrophysics.  The papers reviewed in \cite{pani} focus on possible signatures in black hole collisions generating gravitational waves.  In this context, the ``ringdowns'' following the merger of two holes will have a different structure if there is a true horizon in the larger hole resulting from the merger, or if instead the larger exotic compact object resulting from the merger is bounded by a surface that is not a horizon.    We refer the reader to the papers reviewed in \cite{pani} for further details, which are complex.

Another way in which a horizonless hole could have observational consequences is if it is ``leaky'', that is if interior particles can leak out through its surface \cite{adler2}.  The calculations given above show that  a Weyl scaling invariant dark energy action leads not only to the absence of a horizon in  Schwarzschild-like holes \cite{AR} arising from solving the vacuum Einstein equations, but  leads  also to a small black hole wind  in horizonless holes resulting when matter is present with a pressure-dependent jump in the interior equation of state.  However, this wind is too small to have astrophysical consequences, which is a good thing since observed holes do not evaporate on observational time scales.

In order for a black hole wind to be large enough to account for astrophysical processes such as new star formation near a black hole \cite{adler3} or galaxy formation \cite{adler4}, it will have to arise from the case in which the black hole is accreting infalling matter.  Then, since the interior metric $g_{00}$ never precisely approaches zero, particles entering the hole can get out with a time delay depending on the impact parameter, as noted above.  This could give rise to an exiting wind large enough to have astrophysical consequences.  It could also, because of the possibility of large time delays, offer an explanation of the recently observed two year time delay \cite{delay} between a black hole tidal disruption event in which a star is devoured and the subsequent ejection of some of the absorbed matter.  These possible astrophysical consequences of absence of a horizon, at present speculative,  merit further detailed investigation.

\section{Acknowledgements}

The calculations of this paper were performed with benefit from the hospitality of
the Aspen Center for Physics, which is supported by the National Science Foundation grant PHY-1607611.

\appendix

\section{The continuity argument}
Consider the first order differential equation
\begin{equation}\label{sample1}
\frac{dF(r)}{dr}=G(r)~~~,
\end{equation}
on the domain $r_A \leq  r \leq r_B$.  Then if $|G(r)|$ is bounded by ${\cal B}$ on this domain, the solution $F(r)$ must be continuous. To prove this, pick an arbitrary point $r_0$ in the domain,  integrate Eq. \eqref{sample1} from $r_0-\xi$ to $r_0+\xi$ with $\xi >0$, and take the absolute value, giving
\begin{equation}\label{sample2}
|F(r_0+\xi)-F(r_0-\xi)|=\int_{r_0-\xi}^{r_0+\xi} dr| G(r) |\leq 2 \xi {\cal B}~~~.
\end{equation}
Letting $\xi \to 0$, the right hand side of Eq. \eqref{sample2} vanishes, showing that $F(r)$ is continuous at $r_0$.

\section{Rescaling to dimensionless variables}

In the numerical work it is convenient to rescale to dimensionless variables, with the rescaled central matter  pressure set to unity.  This is accomplished by defining rescaled variables $\bar r= r  p(0)^{1/2}$, $\bar M= M  p(0)^{1/2}$,  $\bar{\hat{m}}=\hat{m}p(0)^{1/2}$,  $\bar p(r)=p(r)/ p(0)$, $\bar \rho(r)= \rho(r)/p(0)$, $\bar \Lambda=\Lambda/ p(0)$, $\bar{\nu}=\nu$, etc.,  which can be verified to be an invariance of the TOV equations.  In the text and program, we exclusively use rescaled variables, with the overbar notation  omitted.

To determine the rescaled value $\bar \Lambda$  corresponding to a given black hole mass $M$, we use the scaling relation
$ \bar \Lambda = M^2 \Lambda/(\bar M)^2$.  Using the observed value  $\Lambda=1.3 \times 10^{-52} {\rm m}^{-2}$, for black hole masses of $M_\odot, \, 10^6 M_\odot, \, 10^8 M_\odot$ one finds for the product $M^2 \Lambda$ the values $0.3 \times 10^{-47}, \, 0.3 \times 10^{-35} , \, 0.3 \times 10^{-31}$ respectively.  Using this, and dividing by the squared rescaled mass values in Table II, one gets the corresponding rescaled cosmological constant values, denoted lambda in the programs, as given in Table II for the example of a mass $10^6 M_\odot$ hole.

\section{Relativistic conserved flux calculation}
The simplest way to find the conserved flux is to note that when $T_{\mu}^{\nu}$ is a covariantly conserved stress energy tensor, $\nabla_\nu
T_{\mu}^{\nu}=0$, then the mixed Einstein--Dirac pseudotensor $t_{\mu}^{\nu}$ \cite{dirac}, when added to $T_{\mu}^{\nu}$,  gives a conserved quantity $\partial_{\nu} \big(({}^{(4)}g)^{1/2}(T_{\mu}^{\nu}+t_{\mu}^{\nu}) \big)=0.$  In a static context, when all quantities are time-independent, the formula given in \cite{dirac} shows that $t_0^r$ vanishes, implying that
\begin{equation}\label{conss1}
(d/dr)\big(({}^{(4)}g)^{1/2} T_0^r\big)=0~~~.
\end{equation}
One way to verify this directly is to use the affine connection formula for the static case
\begin{equation}\label{affine}
\nabla_{\nu}T_0^{\nu}= ({}^{(4)}g)^{-1/2} \partial_j\big(({}^{(4)}g)^{1/2} T_0^j\big) -\Gamma^\kappa_{0\nu}g^{\nu \lambda} T_{\lambda \kappa}~~~,
\end{equation}
and to observe that $A^{\kappa \lambda}= \Gamma^\kappa_{0\nu}g^{\nu \lambda}$  is an antisymmetric tensor. Hence $\nabla_{\nu}T_0^{\nu}=0$ implies $\partial_j\big(({}^{(4)}g)^{1/2} T_0^j\big)=0$.  A second way to verify this directly, for the specific construction used in the wind calculation, is to use Mathematica to show algebraically that the covariant conservation equation $D^{r}=0$ (see Eq. \eqref{ddeff}) is identical to Eq. \eqref{conss1}.  Applying Eq. \eqref{conss1} to Eq. \eqref{enmomgas}, and fixing overall constant factors by requiring the correct nonrelativistic limit, gives Eq. \eqref{consflux}.

\vfill\eject

 \begin{figure}[t]
\begin{centering}
\includegraphics[natwidth=\textwidth,natheight=300,scale=0.8]{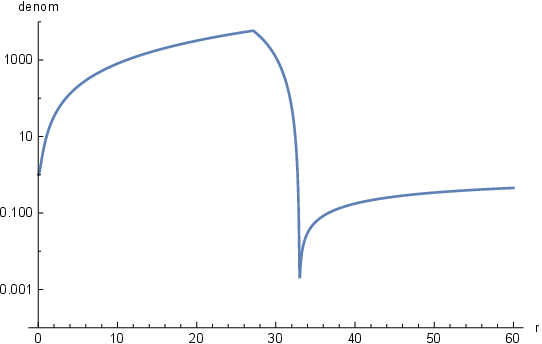}
\caption{TOV denominator denom=${\cal D}$ in the TOV.01 notebook.   }
\end{centering}
\end{figure}

 \begin{figure}[b]
\begin{centering}
\includegraphics[natwidth=\textwidth,natheight=300,scale=0.8]{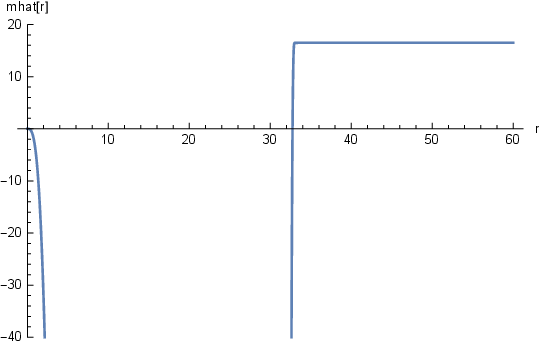}
\caption{$\hat m(r)=1-{\cal D}*r/2$ in the TOV.01 notebook. This levels off at the effective hole mass $M$, at a radius of $2M$.  }
\end{centering}
\end{figure}

 \begin{figure}[t]
\begin{centering}
\includegraphics[natwidth=\textwidth,natheight=300,scale=0.8]{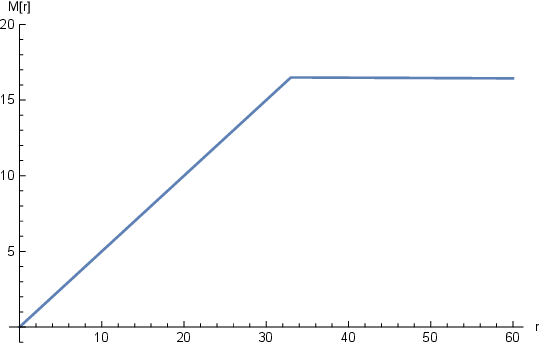}
\caption{$\big(1-\exp(\nu(r)\big)*r/2=  M(r)$ in the TOV.01 notebook.   This levels off at the effective hole mass $M$, at a radius of $2M$.  The initial value $\nu(0)=$nuinit is adjusted to achieve this leveling off.  }
\end{centering}
\end{figure}

 \begin{figure}[b]
\begin{centering}
\includegraphics[natwidth=\textwidth,natheight=300,scale=0.8]{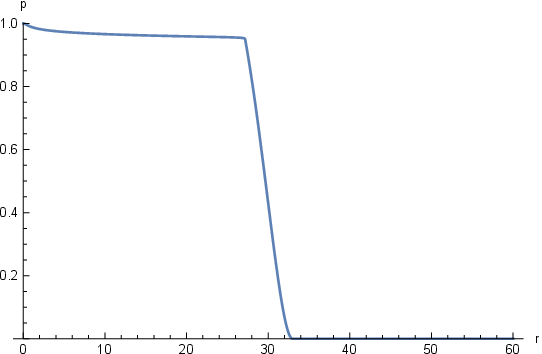}
\caption{$\hat p(r)$ in the TOV.01 notebook. The plot of $p(r)$ looks the same.}
\end{centering}
\end{figure}

 \begin{figure}[t]
\begin{centering}
\includegraphics[natwidth=\textwidth,natheight=300,scale=0.8]{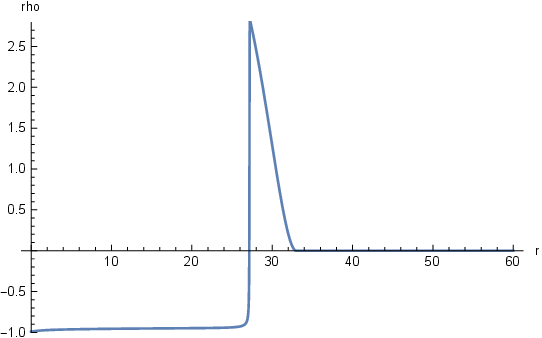}
\caption{$\hat \rho(r)$ in the TOV.01 notebook. The plot of $\rho(r)$ looks the same.}
\end{centering}
\end{figure}

\begin{figure}[b]
\begin{centering}
\includegraphics[natwidth=\textwidth,natheight=300,scale=0.8]{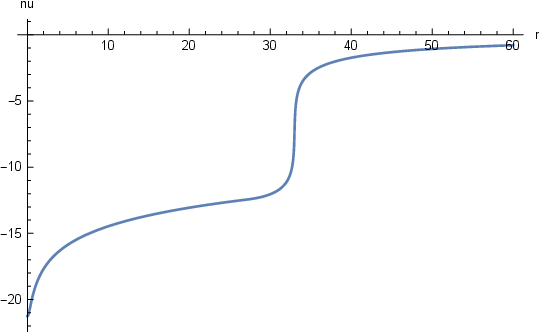}
\caption{ Plot of $\nu(r)$ in the TOV.01 notebook. }
\end{centering}
\end{figure}

 \begin{figure}[t]
\begin{centering}
\includegraphics[natwidth=\textwidth,natheight=300,scale=0.8]{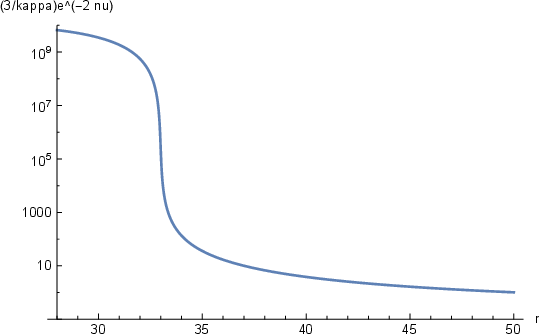}
\caption{Plot of $(3/\kappa) \exp\big(-2\nu(r)\big)$ in the TOV.01 notebook.  The value of this at $r=3M \simeq 49.5$   (see table II), multiplied by $\Lambda$, sets the magnitude of the black hole wind. }
\end{centering}
\end{figure}

 \begin{figure}[b]
\begin{centering}
\includegraphics[natwidth=\textwidth,natheight=300,scale=0.8]{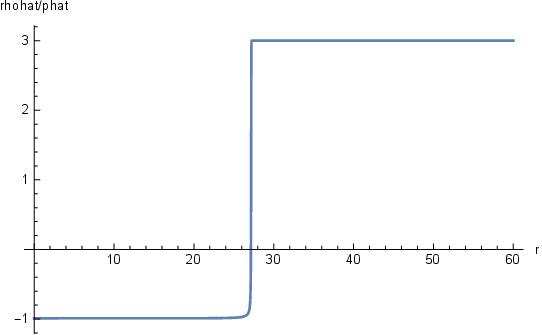}
\caption{Ratio $\hat \rho/\hat p$, essentially the same as $\rho/p$, showing where the equation of state jump occurs.}
\end{centering}
\end{figure}

 \begin{figure}[b]
\begin{centering}
\includegraphics[natwidth=\textwidth,natheight=300,scale=0.8]{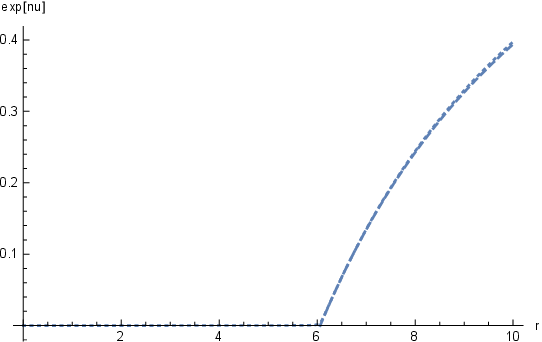}
\caption{Dotted line is $g_{00}=\exp\big(\nu(r)\big)$ for $\beta=.1$; dashed line is $1-6.06/r$, showing they nearly coincide for $r > 6.06 $.}
\end{centering}
\end{figure}

 \begin{figure}[b]
\begin{centering}
\includegraphics[natwidth=\textwidth,natheight=300,scale=0.8]{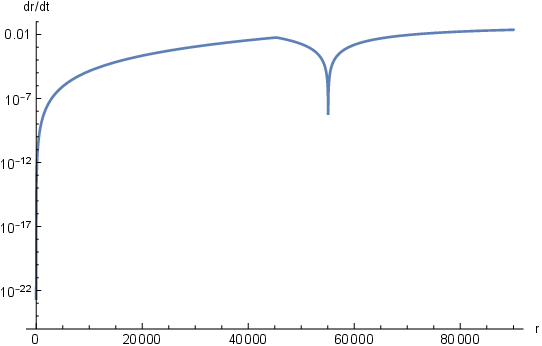}
\caption{$dr/dt$ for a radially infalling particle starting with zero velocity at spatial infinity, for $\beta=.001$. This is an upper bound on $dr/dt$ when the angular momentum per unit mass is nonzero.}
\end{centering}
\end{figure}

\end{document}